\documentclass[10pt,aps,prd,nofootinbib,reprint, superscriptaddress]{revtex4-1}

\usepackage[english]{babel}
\usepackage{tikz}
\usetikzlibrary{angles,quotes}
\usetikzlibrary{calc}
\usetikzlibrary{patterns}

\usepackage{xcolor}
\usepackage{verbatim}

\usepackage{graphicx}
\usepackage{dcolumn}
\usepackage{bm}

\usepackage{pgfplots}
\pgfplotsset{compat=newest}

\pdfoutput=1
\usepackage{url}
\usepackage{hyperref}
\usepackage{hyperref}
\hypersetup{colorlinks=true,allcolors=[rgb]{0.5,0.0,0.5}}
\usepackage{amsfonts}
\usepackage{epsfig}
\usepackage{dcolumn}
\usepackage{xcolor}
\usepackage{amsmath}
\usepackage{amssymb}
\usepackage{amsthm}
\usepackage{amsfonts}
\usepackage{verbatim}
\usepackage{enumerate}
\usepackage{graphicx}
\usepackage{accents}
\usepackage{bm}

\usepackage{physics}
\usepackage{tensor}
\usepackage{siunitx}
\usepackage{wasysym}

\usepackage{graphics}
\usepackage{epstopdf}
\usepackage[utf8]{inputenc} 
\usepackage{pgfplots}
\usepackage{slashed}

\usepackage{braket}
\usepackage{dsfont}
\usepackage{float}
\usepackage[normalem]{ulem}

\newcommand{\ed}[1]{}
\def\vev#1{\langle #1\rangle} 
\def\eq#1{Eq.~(\ref{#1})}

\def\bp{{\vec{\mathbf p}}}

\renewcommand\({\left(}
\renewcommand\){\right)}
\renewcommand\[{\left[}
\renewcommand\]{\right]}

\newcommand\eg{{\it e.g.}~}
\newcommand\ie{{\it i.e.}~}

\newcommand{\del}{\partial}

\newcommand\ee{\end{equation}}
\newcommand\be{\begin{equation}}
\newcommand\eea{\end{eqnarray}}
\newcommand\bea{\begin{eqnarray}}

\begin{document}

\title{On the Tremaine-Gunn Limit with Mass-Varying Particles}

\author{Lotfi Boubekeur}
\email{lotfi.boubekeur@epn.edu.ec}
\affiliation{International Centre for Theoretical Physics\\
Strada Costiera 11, 34151, Trieste, Italy.}
\affiliation{Escuela Polit\'ecnica Nacional, Departamento de F\'isica\\
Ladr\'on de Guevara E11-258, Quito, Ecuador.}
\author{Stefano Profumo}
\email{profumo@ucsc.edu}
\affiliation{Department of Physics and Santa Cruz Institute for Particle Physics\\University of California, Santa Cruz, CA 95064, USA.}

\begin{abstract} General classical arguments on the time evolution of the phase-space density can be used to derive constraints on the mass of particle candidates for the cosmological dark matter (DM). The resulting Tremaine-Gunn limit is  extremely useful in constraining particle DM models. In certain models, however, the DM  particle mass varies appreciably over time. In this work, we generalize the phase-space limits on possible DM particle masses to these scenarios.  We then examine the ensuing cosmological implications on the effective DM  equation of state and indirect DM detection.
\end{abstract}
\date{\today}
\maketitle

\twocolumngrid

\section{Introduction}
    The nature of the cosmological Dark Matter (DM) continues to stand as one of the greatest mysteries at the interface of particle physics and cosmology \cite{Zyla:2020zbs}. There exist conclusive evidence that the solution to  this puzzle lies beyond the standard model of particle physics: The only standard model particles that possess properties adequate for them to be DM candidates are neutrinos; However, neutrinos were quickly discarded  as a DM candidate using phase space arguments \cite{Tremaine:1979we}: In the absence of interactions leading to collision, dissipation, or particle number changing processes, quantum mechanics dictates that any fermionic species possesses a maximal phase-space density $f_{\rm max}$ and that any coarse-grained phase-space density must necessarily be smaller than $f_{\rm max}$. Applying this argument to light neutrinos with $g_\nu$ internal degrees of freedom, and considering observational data from DM-dominated structure, leads to the so-called Tremaine-Gunn limit~\cite{Tremaine:1979we}
\be 
m_\nu\gtrsim 100 {\rm ~eV} \(\frac{4}{g_\nu}\)^{1/4}  \(\frac{100\, \rm{km/s}}{\sigma_v}\)^{1/4}\( \frac{\rm kpc} {r_c} \)^{1/2}, 
\ee 
 where $\sigma_v$ is the structure's velocity dispersion, $r_c=\sqrt{9 \sigma_v^2/4\pi G_N \bar{\rho}}$ is the core (or King) radius,  and in the expression for $r_c$ $\bar{\rho}$ is the core density \cite{Tremaine:1979we,2008gady.bookB}. For instance, applying this limit to the Fornax dwarf spheroidal galaxy (dSphs hereafter), one obtains $m_{\nu}>164$ eV \cite{Boyarsky:2008ju}. Notice, however, that in the presence of interactions, the Tremaine-Gunn limit can be evaded, and lighter thermal DM candidates are observationally viable, see \eg \cite{Alexander:2020wpm}. A second way to evade the  lower mass bound is by postulating a large number of fermionic species with almost degenerate masses, filling up phase space without saturating it \cite{Davoudiasl:2020uig}. A generalization including bosons (and fermions) with non-vanishing chemical potential was explored in a series of subsequent papers \cite{Madsen:1990pe,
Madsen:1991mz, Madsen:2000zd}. Since the primordial DM phase-space density is not accessible through astrophysical observations, a surrogate quantity, more suitable for observational purposes, was defined in Ref.~\cite{Hogan:2000bv} as  
\be 
Q\equiv\frac{\bar{\rho}}{\vev{v^2}^{3/2}}= \frac{\bar{\rho}}{3\sigma_v^3}\,.
\label{q}
\ee
In \eq{q}, the quantities on the right-hand side are determined from observations. Like its primordial counterpart, the average halo phase density $Q$ cannot increase with time \cite{Hogan:2000bv, Dalcanton:2000hn} during the evolution of DM in halos {\it i.e.} $Q_i>Q_f$. This behavior has been proven to persist even after significant phase mixing and violent relaxation phases through explicit $N$-body simulations \cite{Peirani:2005kw, 2007astro.ph..1292P}. This property yields different, but related, phase-space arguments, that when applied to dSphs lead, in turn, to lower bounds on DM particles \cite{Hogan:2000bv,Dalcanton:2000hn, Boyanovsky:2007ay, Boyarsky:2008ju, Gorbunov:2008ka} which are quite comparable to the original Tremaine-Gunn limit. For recent works along these lines, see \eg \cite{2013MNRAS.430.2346S, 2021MNRAS.501.1188A,Domcke:2014kla,Randall:2016bqw,DiPaolo:2017geq,Giraud:2018gxl, Savchenko:2019qnn}

The  Tremaine-Gunn limit is a powerful constraint to constrain the parameter space of particle DM models. The purpose of this note is to extend the use of phase-space arguments of the kind that lead to the Tremaine-Gunn  limit to more general situations where the mass of the DM particle evolves with time or scale factor, \ie $m(a)$ with $a$ the cosmological scale factor. Since fermionic DM fields generally get their masses through vacuum expectation values (VEV) of scalar fields $\phi_i$  (Yukawa couplings), it is natural to expect that DM masses are dynamical parameters controlled by the the dynamics of scalars,  $m_{\rm dm}\propto \phi_i(t)$. For instance, Ref.~\cite{Berlin:2016bdv}, aimed to decouple the constraints stemming from sterile neutrino radiative decay by allowing the mass to be a dynamical parameter {\it i.e.} $m=m(a)$. In this scenario, neutrino masses are controlled by the vacuum expectation value (VEV) of a pseudoscalar field (an axion or an axion-like particle). This is only but an instance of a scenario where the DM particle mass varies considerably over the course of the evolution of the universe (See for instance  \cite{Casas:1991ky,  Garcia-Bellido:1992xlz,   Anderson:1997un, Franca:2003zg, Rosenfeld:2005pw} for earlier works and \cite{Davoudiasl:2019xeb} for more recent studies).

Motivated by this, in this note we reconsider phase-space constraints on the DM particle mass in scenarios where the latter evolves with the scale factor. We explore the consequences of a variable mass for cosmological observations, and outline new constraints from gamma-ray observations.

Throughout this paper we use natural units: $\hbar=c=k_B=1$. The remainder of the paper is organized as follows. In Section \ref{sec2}, we derive the constraints on the mass variation of a DM particle using general classical arguments on the evolution of phase-space. In Section \ref{sec3}, we consider the cosmological consequences and bounds on the scenario. In Section \ref{sec4}, we consider yet another signature of the scenario that leads to significantly modified expectations for gamma rays from DM annihilations at all redshifts. Finally in Section \ref{sec:conclusions}, we give our conclusions and outlook. 

\section{The Liouville theorem}
\label{sec2}
The Liouville theorem, describing the evolution of a collisionless and dissipationless system  of particles in phase space, is a cornerstone of classical mechanics \cite{Goldstein}. The theorem states that under certain circumstances the volume of phase space occupied by a set of $N$ particles remains {\em constant} over time. The theorem hinges on the following two assumptions: (i) the total number of particles in phase space is constant, i.e. $dN/dt=0$, and (ii) the phase space density, in absence of interaction,  obeys the identity 
\bea
\label{boltz}
\frac{d f (\vec{\mathbf p}, \vec{\mathbf q})}{d t}=\frac{\del f}{\del t}+ \frac{d \vec{\mathbf q}}{d t}\cdot \vec{\nabla}_{\vec{\mathbf q}} f + \frac{d \vec{\mathbf p}}{d t}\cdot \vec{\nabla}_{\vec{\mathbf p}} f=0\,.
%
%
%
\eea
By writing $N=\int_{\textrm{Phase space}} f(\vec{\mathbf p}, \vec{\mathbf q})\,\,  {d \vec{\mathbf p}}\, {d\vec{\mathbf q}}\,$ 
and combining with assumptions (i) and (ii) above, one can prove readily that 
\be
\frac{d}{dt}\int_{\textrm{Phase space}} f(\vec{\mathbf p}, \vec{\mathbf q})\,  {d \vec{\mathbf p}}\, {d\vec{\mathbf q}} =0\,.
\ee
In what follows, and in order to apply the Liouville theorem, we will assume that (i) the particles under investigation (whether they decoupled while relativistic or not) have already {\em chemically} decoupled, so that {\em number-changing} processes can effectively be neglected,  but that the particles are  still in {\em kinetic} equilibrium.

A very useful and frequently encountered situation occurs when the phase-space density depends only on the  Hamiltonian (i.e. there is no explicit time-dependence:  $\del f(H)/\del t=0$). In this case  \eq{boltz}
can be cast as 
\bea
\frac{d}{d t} f (H)=
\frac{\del f}{\del H}\sum_i\[\frac{\del H}{\del p_i}\dot{p_i} + \frac{\del H}{\del q_i}\dot{q}_i\] =0\, ,
\eea
 automatically satisfying Liouville's theorem. This happens often in equilibrium statistical mechanics where energy is conserved. Note, however, that this argument does not depend on whether energy is conserved or not (i.e. whether the Hamiltonian is time-dependent or not). 

In general relativity, the phase-space density obeys $\hat{L}[f]=\hat{C}[f]$, where $\hat{L}$ is the usual Liouville operator given by  
\be 
\hat{L}=p^\alpha \del_\alpha -\Gamma^\alpha_{\beta\gamma} p^\beta p^\gamma\frac{\del}{\del p^\alpha}\,, 
\ee
and $\hat{C}$ is the collision operator, describing scatterings, production and annihilation (processes that would change the number of particles). Considering a collisionless and dissipationless system of particles in a geometry described by a Friedmann-Robertson-Walker (FRW) metric, one gets \footnote{See \eg  \cite{Kolb:1990vq}, Eq. (5.5).}
\be 
E \frac{\del f}{\del t}-H|\bp|^2 \frac{\del f}{\del E}=0
\label{liouv}
\ee
It is instructive to see how this last identity \eq{liouv} is satisfied explicitly: in an isotropic and homogeneous universe, the phase-space density of a species that {\em chemically} decoupled at at temperature $T_{\rm dec}$, when the scale factor was $a_{\rm dec}$, is given by \cite{Kolb:1990vq} 
 \be 
f(|\bp|, T)=f\(|\bp_{\rm dec}| a_{\rm dec} / a, T\)\,, 
\ee
where $a(t)$ is the scale factor, and where we used the usual redshifting relation for momenta. This scaling will be valid as long as kinetic equilibrium is maintained. Now, to go further, we need to specify the relationship between temperature and scale factor for hot and cold relics separately \cite{Kolb:1990vq}
\be
T\sim\left\{\begin{array}{cc} T_{\rm dec}\, \displaystyle\cdot \(\frac{a_{\rm dec}}{a}\) & \textrm{for  }\,\, T_{\rm dec}\gg m \\
~ & ~ \\
T_{\rm dec}\, \displaystyle\cdot \(\frac{a_{\rm dec}}{a}\)^2  & \textrm{for  }\,\, T_{\rm dec}\ll m\\
\end{array}\right.
\ee
This means that the phase space density will remain constant in both cases, since for $T_{\rm dec}\gg m$ the energy $E\simeq |\vec{p}|$ while for $T_{\rm dec}\ll m$, $E\simeq |\vec{p}|^2/(2m)$.  More explicitly: 
\be 
f(|\bp|, T)=\frac{1}{e^{E/T}\pm 1}= \frac{1}{e^{E_{\rm dec}/T_{\rm dec}}\pm 1}\,,  
\label{fg}
\ee 
where the sign $+$($-$) is for fermions (bosons). It follows that, after decoupling the phase space density is constant $df(E,  T)/dt=0$, in agreement with the Liouville theorem. Note that this is not the case of a species that decouples while semi-relativistic, i.e. when $T_{\rm dec}\sim m$: in that case the phase space distribution does not maintain the equilibrium form absent interactions \cite{Kolb:1990vq}. In any case, if the particle under consideration is to be a significant fraction of the cosmological dark matter, structure formation enforces it to decouple when non-relativistic, hence the assumption that $df(E,  T)/dt=0$ applies.

The Tremaine-Gunn limit states that the coarse-grained space phase density is bounded from above by the maximum value of its fine-grained counterpart. Let us now calculate 
\bea 
\frac{df}{dT}\Big|_E&=&\frac{1}{\dot{T}(t)}\frac{\del f}{\del t}=- \frac{|\bp|^2}{T E} \frac{\del f}{\del E} \nonumber\\&=&
\left\{\begin{array}{cc} \frac{|\bp|^2}{T^2 E} \frac{e^{E/T}}{(e^{E/T}\pm 1)^2}\,

& \textrm{for  }\,\, T_{\rm dec}\gg m \\
~ & ~ \\
\frac{|\bp|^2}{2 T^2 E} \frac{e^{E/T}}{(e^{E/T}\pm 1)^2} & \textrm{for  }\,\, T_{\rm dec}\ll m\\
\end{array}\right.
\eea
where we have used \eq{liouv} and the fine-grained phase space density \eq{fg}.  This means that the phase space density  {\it  monotonically} decrease with time, and thus increase with temperature in a FRW universe {\it i.e}
\be 
f(E, T_1)\ge f(E, T_2), \quad {\rm for~} T_1 \ge T_2\,.
\label{f}
\ee
Now, following the line of reasoning of \cite{Tremaine:1979we}, we will apply \eq{f} in its coarse-grained version to a fermionic \footnote{We will focus on the case of a fermionic DM particle with a vanishing chemical potential $\mu=0$, as in the original Tremaine-Gunn limit. The inclusion of non-vanishing chemical potentials has been considered in \cite{Madsen:1990pe, Madsen:1991mz, Madsen:2000zd}} DM halo with an evolving DM particle mass. As in the original Tremaine-Gunn limit, we will adopt the simplifying assumption that DM particles will collapse to a self-gravitating system  described by an isothermal sphere \cite{2008gady.bookB} with velocities following a Maxwell-Boltzmann distribution
\be 
f_{\rm halo}(\vec{\mathbf{p}}, r)\, d \vec{\mathbf{p}} =\frac{n_{\rm halo}(r)}{(2\pi\sigma_v^2)^{3/2} m^3} \, e^{-{v^2}/{2\sigma^2_{v}}}\, d \vec{\mathbf{p}},
\label{fhalo}
\ee
where $\sigma_v=\sqrt{T/m}$ is the Maxwellian velocity dispersion at any given temperature and $n_{\rm halo}(r)$ is the density of DM in the halo  
\be 
n_{\rm halo}(r)=\frac{\sigma^2_v}{2\pi G_N r^2 m}\,.
\label{nhalo}
\ee
In the strict sense, our results are correct for an isotropic halo. A more accurate description of the halo would include a proper Jeans analysis, taking into account the anisotropy parameter $\beta(r)=1-\vev{v^2_\theta}/\vev{v^2_r}$. However, the effect of $\beta(r)$ is to rescale the phase-space density with a factor $\[2\log\(r_t/ r_c\)\]^{-3/2}$, where $r_t$ is the tidal radius \cite{Gerhard_Spergel}. Therefore, we do not expect this logarithmic correction to affect significantly our analysis.

Since $df_{\rm halo}/dt=0$, by virtue of the Liouville theorem, this implies that  $n_{\rm halo}(r)=\int f_{\rm halo} \,d \vec{\mathbf{p}}$ is also a constant {\it i.e. } $d n_{\rm halo}/dt=0$. Following \eq{f}, we get $f_{\rm{halo}}(T_1)\Big{|}_{max}\ge f_{\rm halo}(T_2)\Big{|}_{max}$, which in turn implies
\be
\frac{m(T_2)}{m(T_1)}\ge \frac{T_1}{T_2}>1\,.
\label{eq17}
\ee
This is the main result of this note: {\em the DM mass change in the halo can only increase with time, as a consequence of phase-space density limits}. Note that the inequality \eq{eq17}, constraining the mass to monotonically increase with time, was obtained using the phase-space density $f_{\rm halo}(\vec{\mathbf{p}}, r)$. We can obtain the same inequality using the surrogate quantity $Q$ defined in \eq{q}.  As noted in \cite{Boyarsky:2008ju, Gorbunov:2008ka}, the average halo phase-space density scales as $Q\sim m^4 f_{\rm halo}\,.$  On the other hand, $f_{\rm halo}$ scales as $m^{-4}$, see \eq{fhalo} and \eq{nhalo}. This means that one will obtain the inequality \eq{eq17} also as a consequence of $Q_i>Q_f$.

Let us now comment on this important result. First,  Eq.~\ref{eq17} is consistent with expectations from general thermodynamical arguments applied to an isothermal DM-dominated halo in hydrostatic equilibrium at constant volume and temperature. To prove this, notice that since the number of DM particles in the halo $N=4\pi\int n_{\rm halo}(r) r^2 dr$ is conserved, this yields the scaling $\sigma^2_v\sim m$, or $T =  m^2/C$, where $C$ is positive. Next, combining the hydrostatic equilibrium equation of the halo \be 
\frac{dp}{dr} =-\frac{G_N M(r)}{r^2}\rho=\sigma^2_v \frac{d\rho}{dr}\,,
\ee
with the first law of thermodynamics one gets
\be 
dS=(1+\sigma^2_v) \rho\, dV + N\frac{dm}{T}\,. 
\ee
This in turn allows us to compute the entropy change between $t_1$ and $t_2$:  
\bea
\Delta S_{12}&=&\int_{m(t_1)}^{m(t_2)}\(\frac{\del S}{\del m} \)_V  dm =
\int_{m(t_1)}^{m(t_2)} N \frac{dm}{T} \nonumber\\&=& 
C N \( \frac{1}{m(t_1)} - \frac{1}{m(t_2)}\) \,.
\label{s}
\eea
 The second principle of thermodynamics implies \eq{s} should be positive, therefore $m(t_2)>m(t_1)$. Note that this inequality here assumes that (i) halos are in hydrostatic equilibrium, and that (ii) $T$ and $V$ do not change.

The second comment is that inequality \eq{eq17} was derived without assuming that the DM particles were created thermally, or were in equilibrium before collapsing into halos described by Maxwellian velocity distribution. This means that it can applied quite generally to mass-varying DM scenarios irrespective of their production mechanism.

\section{Cosmological consequences and bounds}
\label{sec3}
Let us now discuss the main cosmological consequences of having a cold DM particle with a monotonically increasing time-dependent mass $m_{\rm dm}(a)$. We will focus on the impact of mass variation on the DM equation of state (EoS) $\omega_{\rm dm}=p_{\rm dm}/\rho_{\rm dm}$. As we will see shortly, we can recast the experimental bounds on $\omega_{\rm dm}$ as bounds on the DM mass variation.  In the standard case $m'_{\rm dm}(a)=0$, and after DM decoupling $T<T_{\rm dec}$ the value of the equation of state parameter is 
\be 
\omega^{(0)}_{\rm dm}=\frac{T_{\rm dec}}{m_{\rm dm}}\cdot\(\frac{a_{\rm dec}}{a}\)^2=10^{-11}\(\frac{T_{\rm dec}}{\rm eV} \) \(\frac{{100\, \rm GeV}}{m_{\rm dm}} \)\cdot\(\frac{a_{\rm dec}}{a}\)^2\,.  
\label{w0}
\ee
This is an extremely small number which justifies the approximation $\omega_{\rm dm}=0$ at all times. Now, if the mass of the DM particle varies, while retaining its stability, the number density of DM will obey $\dot{n}_{\rm dm}+ 3 H n_{\rm dm}=0$ exactly as in the standard case $m'_{\rm}(a)=0$. However, the DM energy density $\rho_{\rm dm}=m_{\rm dm}(a)\cdot n_{\rm dm}$ will obey a  different continuity equation which can be seen as an effective DM equation of state $\dot{\rho}_{\rm dm}+ 3 H \rho_{\rm dm}(1+\omega_{\rm dm})=0$, with 
\be 
w_{\rm dm}=-\frac13 \frac{\del \log m_{\rm dm}(a)}{\del \log a}\,.
\label{wdm}
\ee


Since, as we have concluded in section \ref{sec2},  $m_{\rm dm}$ grows monotonically with the scale factor, \eq{wdm} means that the effective DM EoS is {\it negative} \ie  $\omega_{\rm dm}\le 0$. We can also generalize this result to include scenarios where only a fraction $f\equiv\rho_{\rm mv}/\rho_{\rm tot}=$ of the total DM has $m_{\rm dm}'(a)\ne 0$.  In this case, \eq{wdm} becomes 
\be 
w_{\rm dm}=-\frac{f}{3} \frac{\del \log m_{\rm dm}(a)}{\del \log a}\le 0\,.
\label{ff}
\ee

\begin{figure}[!t]
    \centering
\vspace{1 cm}
   \includegraphics[width=0.5\textwidth]{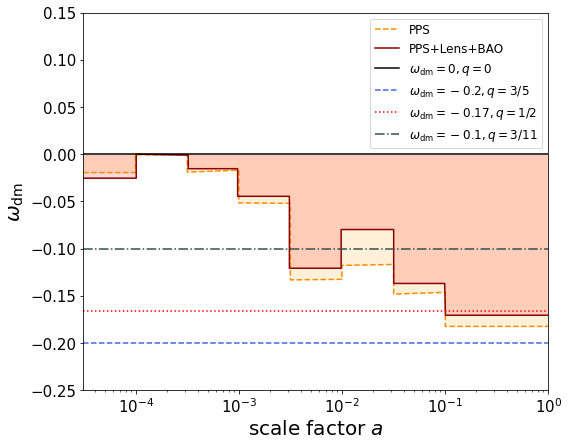} 
   \caption{Constraints (99$\%$ CL contours) on $\omega_{\rm dm}$ from \cite{Ilic:2020onu} using Planck Power Spectrum (PPS) and BAO and Lensing data. The different datasets combinations are specified in the legend. The horizontal lines stand for different varying DM masses, where $m(a)= m_0(a/a_0)^q$.}
   \label{fig:omega}
\end{figure}

This is the second main result of this note. The contribution \eq{ff} to the DM EoS is expected to dominate over the standard one, and as such we will neglect \eq{w0} with respect to \eq{ff}.  In the following,  we will explore its observational constraints. As we have seen,  \eq{f} means that we can rephrase any bound on the variation of $m_{\rm dm}(a)$ using the existing bounds on $\omega_{\rm dm}$.  This last quantity describes how cold/warm is the DM particle, and as such it has been the subject of multitude of studies with a variety of motivations and emphasis \cite{Armendariz-Picon:2013jej, Tutusaus:2016kyl, Serra:2011jh, Faber:2005xc, Kunz:2016yqy, Calabrese:2009zza, Ballesteros:2020adh, Kumar:2012gr, Kumar:2019gfl,  Xu:2013mqe, Avelino:2012jy, Thomas:2016iav, Ilic:2020onu, Kopp:2018zxp, Muller:2004yb}. It has also been investigated in connection to DM bulk viscosity (see \eg \cite{Velten:2012uv}).  In these studies, several datasets, probing different redshift intervals, has been leveraged to extract bound on the equation of state parameter (EoS) parameter of dark matter $\omega_{DM}$. Unfortunately, a large number of the aforementioned studies did not include negative values of $\omega_{DM}$ in their priors (See for instance \cite{Kumar:2019gfl}). The main motivation behind that bias towards positive $\omega_{\rm dm}$ is that one expects a positive corrections to the DM EoS from kinetic pressure. Yet, it has been noted in many studies that negative values of $\omega_{\rm dm}$ occur naturally 
\cite{Calabrese:2009zza, Muller:2004yb, Xu:2013mqe}, and might even alleviate some of the problems/issues of $\Lambda$CDM (See \eg  \cite{Naidoo:2022rda} for a recent study addressing both the $H_0$ and $\sigma_8$ tension).  
For the purpose of this study, we will use the results of the recent study \cite{Ilic:2020onu}, spanning a broad range of redshifts and including negative values of the EoS parameter. The constraints in \cite{Ilic:2020onu} were obtained using the Planck Power Spectrum (PPS), Lensing (Lens) and the Baryon Acoustic Oscillations (BAO).   We plot  resulting bounds\footnote{For the purpose of the present study, we will focus on $\omega_{\rm dm}\le 0$ and null speed of sound and shear viscosity $c_s^2=c^2_{\rm vis}=0$.} in Fig.~(\ref{fig:omega}) where the light yellow (red) area corresponds to the 99$\%$ CL contours of allowed values of $\omega_{\rm dm}$ versus the scale factor $a=1/1+z$ using PPS (PPS+Lens+BAO).  For illustration, we plot horizontal lines (see Figure (\ref{fig:omega})legend) corresponding to the simple power-law model (See \cite{Anderson:1997un} for details) $m_{\rm dm}=m_0 (a/a_0)^q$ and $\omega_{\rm dm}=-q/3$. Here $a_0$ is the scale factor at which $m_{\rm dm}$ starts to evolve. In Fig.~(\ref{fig:omega}), there is no significant difference in using the different datasets combinations. However, and as previously noted in \cite{Ilic:2020onu}, the strongest constraints on $\omega_{\rm dm}$ are around matter-radiation equality $a_{\rm eq}\simeq 3\times 10^{-4}$. At later times $z\simeq 1-10$, the constraints are less stringent and values of $\omega_{\rm dm}\simeq -0.15$ can be attained. Notice that all our bounds can be rescaled through \eq{ff} if only a fraction of DM has $m'_{\rm dm}(a)\ne0$. This leaves more room for specific models where $m_{\rm dm}$ increases at low redshift. Finally, it would be interesting to include more datasets to confirm this conclusion.

  \onecolumngrid

\begin{figure}[!t]
   \centering
   \includegraphics[width=0.45
\textwidth]{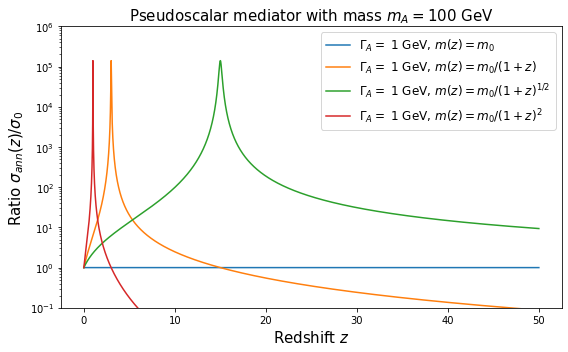} \qquad\includegraphics[width=0.45\textwidth]{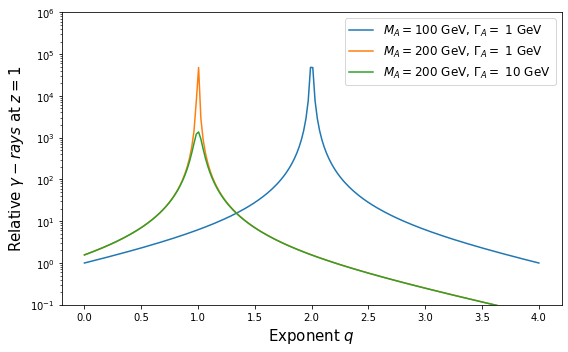}
    \caption{Enhancement of the gamma-ray flux for mass-varying DM masses, where $m(a)=m_0 a^q$. The left panel shows the relative gamma-ray flux at different redshifts for different values of $q$, while the right panel shows the relative gamma-ray flux as a function of $q$ at redshift $z=1$. }
    \label{fig:gammarays}
\end{figure}
\twocolumngrid
\section{Gamma-ray fluxes from mass-varying DM annihilations}
\label{sec4}

In this section, we discuss the impact of DM particle mass variation on the expected flux of gamma rays from its annihilation in dense halos. Neglecting the highly model-dependent effect of mass on the gamma rays produced in a single annihilation event, the flux of gamma rays from DM annihilation in a specific DM halo varies with the DM mass as $\phi_\gamma\sim m(a_0)^{-2}$, where $a_0=\frac{1}{1+z_0},$ and $z_0$ is the DM halo's redshift. However, when considering the signal from a given direction and summing over halos at all distance along that line of sight \cite{Ando:2005hr}, the resulting gamma-ray flux depends quite sensitively on $m_{\rm dm}(a)$:
\bea
 &\phi_\gamma&=\frac{2}{4\pi H_0}\frac{\Omega_\chi^2\rho_{\rm crit}^2}{2}\nonumber \\  &\times&\int dz\frac{(1+z)^3}{h(z)}\frac{\langle\sigma v\rangle(z)}{[m_{\rm dm}(z)]^2}\frac{dN_\gamma(E^\prime_\gamma)}{dE^\prime_\gamma}f(z)e^{-\tau(z,E_\gamma)},
\eea
where $h(z)=[(1+z)^3\Omega_m+\Omega_\Lambda]^{1/2}$,  $E^\prime_\gamma=(1+z)E_\gamma$,  $\tau$ the optical depth, and $f(z)$ a boost factor associated to the ``lumpiness'' of the DM density distribution \cite{Ullio:2002pj,Taylor:2002zd}. Notice that the pair annihilation rate can inherit a redshift dependence from the dependence on redshift of $m_{\rm dm}(z)$. For concreteness, we consider a cross section corresponding to a fermionic DM candidate pair-annihilating to a fermion-antifermion pair via a pseudoscalar mediator of mass $M_A$ and width $\Gamma_A$; the cross section at low velocity depends on the DM mass $m_{\rm dm}(z)$ via the following scaling:
\begin{equation}
\sigma_{\rm ann}(z)\propto\frac{m_{\rm dm}^2(z)}{\left[M_A^2-4m_{\rm dm}^2(z)\right]^2+\Gamma_A^2M_A^2}\,.
\end{equation}

Using the density contrast model ``Bullock et al'' described in detail in \cite{Ullio:2002pj} (see the solid line in Fig.~5, left, of Ref.~\cite{Ullio:2002pj}), and assuming, for simplicity,  negligible effects from the gamma-ray optical depth (corresponding to focusing on low energy photons only) and from the variation of the photon number produced by annihilation at different masses, we find the results shown in Fig.~\ref{fig:gammarays} for the relative gamma-ray flux for a mass-varying $m_{\rm dm}(z)$ versus the constant $m_{\rm dm}(z)=m(0)$ case, for three different choices of the redshift dependence $q$, where $m_{\rm dm}(z)=m_0/(1+z)^q$. The left panel shows the redshift-dependent relative ratio of the mass-varying to constant mass photon flux, for $q=0.5,\ 1$ and $2$; the right panel shows the resulting enhancement (and in some cases suppression, due to the proximity of the on-shell condition for  resonant annihilation) to the photon flux, for $m_A=100$ GeV and $\Gamma_A=1$ GeV (blue line), $m_A=200$ GeV and $\Gamma_A=1$ GeV (orange line) and $m_A=100$ GeV and $\Gamma_A=10$ GeV (green line).

\section{Summary and conclusions}\label{sec:conclusions}
We generalized the Tremaine-Gunn limit on the mass of fermionic dark matter particles to mass-varying dark matter candidates. Assuming persistent kinetic equilibrium, and chemical decoupling as hot or cold relics, the latter being required if the particle under consideration constitutes a significant fraction of cosmological dark matter, we showed that phase-space considerations force the mass to be a growing function of time/scale factor (or equivalently, in turn, a decreasing function of the universe's temperature). We showed that this is consistent with the second principle of thermodynamics as applied to the growing entropy in dark matter halos with time. We showed that the  effective dark matter equation of state resulting from a mass-increasing dark matter candidate can be constrained with observations. Finally, we showed that a generic expectation of a consistent mass-varying dark matter scenario is an increase of indirect dark matter detection rates resulting from dark matter annihilation at all redshifts.

\acknowledgments
SP is partly supported  by the U.S. Department of Energy grant number DE-SC0010107. LB acknowledges the support and hospitality of the HECAP section of ICTP as well as its staff members during the inception of this work. LB is also supported through the ICTP Senior Associateship programme (2023-2028). 
\bibliography{main}
\end{document}